\documentclass[11pt]{article}
\usepackage[utf8]{inputenc}
\usepackage[T1]{fontenc}
\usepackage{graphicx}
\usepackage{dcolumn}
\usepackage{bm}
\usepackage{chngpage}
\usepackage{braket}
\usepackage[margin=0.9in]{geometry}
\usepackage{amsmath}
\usepackage{soul}
\usepackage [english]{babel}
\usepackage [autostyle, english = american]{csquotes}
\MakeOuterQuote{"}
\usepackage{float}
\restylefloat{table}
\usepackage{placeins}
\usepackage{caption}
\usepackage{amsmath, amssymb}
\usepackage{bbold}
\usepackage{makecell}
\usepackage{braket}
\usepackage{calc}
\usepackage{subcaption}
\usepackage{authblk}
\usepackage{natbib}
\usepackage{indentfirst}
\usepackage[labelfont=bf]{caption}
\usepackage[dvipsnames]{xcolor}
\usepackage{hyperref}
\hypersetup{
    colorlinks,
    citecolor=blue,
    filecolor=black,
    linkcolor=blue,
    urlcolor=black
}

\DeclareUnicodeCharacter{2009}{\,} 
\usepackage{tabularx, booktabs}
\usepackage{multirow}
\usepackage{tikz}
\usepackage[left]{lineno}

\usepackage{geometry}
\usepackage{lastpage}
\usepackage{fancyhdr}
\newcolumntype{m}{>{\hsize=.24\hsize \centering\arraybackslash\arraybackslash}X}
\newcolumntype{n}{>{\hsize=.18\hsize \centering\arraybackslash\arraybackslash}X}
\newcolumntype{l}{>{\hsize=.18\hsize}X}
\newcolumntype{o}{>{\hsize=.15\hsize \centering\arraybackslash\arraybackslash}X}

\title{\textbf{Efficient high-harmonic generation in van der Waals ferroelectric NbOI$_2$ crystals}}

\author[1, *, $\dag$]{Tianchen Hu}

\author[2, *]{Feng Li}

\author[1]{Junhan Huang}
\author[3]{Chen Qian}
\author[1]{Ruoxuan Ding}
\author[1]{Hao Wang}
\author[1]{Qiaomei Liu}
\author[1]{Qiong Wu}
\author[3]{Ruifeng Lu}
\author[4, $\dag$]{Chunmei Zhang}
\author[5, 1, 4, $\dag$]{Nanlin Wang}

\affil[1]{International Center for Quantum Materials, School of Physics, Peking University, Beijing 100871, China.}
\affil[2]{School of Science, Nanjing University of Posts and Telecommunications, Nanjing 210046, China.}
\affil[3]{Institute of Ultrafast Optical Physics, Department of Applied Physics, Nanjing University of Science and Technology, Nanjing 210094, China.}
\affil[4]{Beijing Academy of Quantum Information Sciences, Beijing 100193, China.}
\affil[5]{Tsung-Dao Lee Institute, Shanghai Jiao Tong University, Shanghai 200240, China.}
\affil[$\dag$]{e-mail: \texttt{tchu@pku.edu.cn}

\texttt{zhangcm@baqis.ac.cn}

\texttt{nlwang@sjtu.edu.cn}}
\affil[*]{These authors contributed equally to this work.}
\date{}

\begin{document}
\maketitle

\section*{Abstract}

Layered NbOX$_2$ ($X=\mathrm{Cl,\,Br,\,I}$), a member of the van der Waals ferroelectric family, exhibits intrinsic ferroelectricity and pronounced nonlinear optical responses, making it a promising candidate for integrated nanophotonics applications. While previous studies have emphasized the material's strong second-order nonlinear responses, higher-order nonlinear responses are still mostly unexplored. This work systematically investigates NbOI$_2$ using high harmonic generation (HHG) spectroscopy. Driven by an intense mid-infrared laser field centered at $\sim4~\mu\mathrm{m}$ wavelength, highly anisotropic odd- and even-order harmonics up to the 16th order are generated at a low peak intensity of $0.4~\mathrm{TW\,cm^{-2}}$, extending beyond the material's bandgap. Both bulk and flake forms of NbOI$_2$ display pronounced harmonic emission from the near-infrared to the deep-ultraviolet spectral region, with a notably high overall conversion efficiency compared to other known materials. Polarization-resolved measurements reveal that even-order harmonics remain aligned with the crystal polar axis regardless of the driving-field orientation, whereas odd-order harmonics are dynamically affected. First-principles calculations suggest that the flat valence band associated with Peierls dimerization enhances HHG efficiency through electron correlation. These findings provide fresh perspectives on HHG in van der Waals ferroelectric materials and facilitate the development of compact and tunable quantum light sources.

\section*{Introduction}

High-harmonic generation (HHG)\cite{Corkum_1993,Lewenstein1994} is an extreme nonlinear optical process that has been extensively investigated in atomic and molecular gases, enabling the generation of coherent broadband radiation in the extreme-ultraviolet regime and providing a direct route to attosecond pulse production \cite{Popmintchev_2010, Popmintchev_2012}. In recent years, HHG in solids\cite{Ghimire_2010} has been extended to a broad class of condensed-matter systems, including monolayer transition-metal dichalcogenides\cite{Liu2017,Shi2023}, topological materials\cite{ Yoshikawa2017,Bai_2020,Kovalev2020,Schmid2021,Lv2021,Heide2022}, strongly correlated systems\cite{PhysRevLett.128.127401,Silva2018,Alcala2022}, and organic crystals\cite{Wiechmann2025}. HHG spectroscopy is intrinsically sensitive to crystal symmetry and band structure, thereby providing a powerful probe of ultrafast dynamics in solids\cite{Silva_2018,Goulielmakis_2022,Heide2024}. Importantly, owing to the much higher electron densities and rich band-structure effects, solid-state HHG can be driven at substantially lower laser intensities than in gases, highlighting its potential for highly efficient and compact attosecond light sources\cite{Vampa_2015,Ghimire-2018}. Although recent efforts to enhance HHG efficiency via nanostructure design are often constrained by available material platforms\cite{Liu2018,Yang2019,Shcherbakov2021}, van der Waals materials stand out due to their strong quantum confinement, relaxed phase-matching requirements, high tunability, and notably large nonlinear optical coefficients, making them highly promising for efficient HHG\cite{Liu2017,Shi2023,Heide_2023,Ma2025,Mushtaq2024}. Furthermore, in integrated photonic systems and on-chip nonlinear devices\cite{Lyu2025,Lin2026}, strict limitations on device dimensions make it particularly crucial to identify layered two-dimensional materials capable of high nonlinear conversion efficiencies\cite{Mushtaq2024,Lyu2025,Vampa_2017}.

As a newly emerged family of van der Waals ferroelectric materials, NbOX$_2$ (X = Cl, Br, I) features a uniquely in-plane polar structure with pronounced anisotropy in its electronic and optical properties\cite{Fang2021,Mortazavi2022,Abdelwahab2022,Guo2024,Huang2025,Ahsanullah2026}. In addition, the Peierls dimerization of NbOX$_2$ generates a correlated flat dispersion in the top of the valence band\cite{Ye2023,https://doi.org/10.48550/arxiv.2510.15080}. The weak band dispersion results in a high density of states, making interband electronic transitions particularly efficient. Recently, NbOX$_2$ has garnered significant attention for its pronounced and tunable second-order nonlinear optical responses in the perturbative regime, including giant second-harmonic generation, colossal terahertz (THz) emission, spontaneous parametric down-conversion, and the efficient bulk photovoltaic effect\cite{Fang2021,Abdelwahab2022,Handa2025,Subedi2025,Guo2023,Feng2025}. These nonlinear optical responses are comparable to those of traditional bulk electro-optic crystals, yet the material is orders of magnitude thinner, making it highly promising for nanophotonics applications. Moreover, recent THz emission studies suggest the presence of ferrons in NbOX$_2$, making the system even more intriguing\cite{Choe2025,Zhang2025}. However, the behavior of NbOX$_2$ under extremely strong-field excitation remains largely unexplored. Consequently, performing HHG spectroscopy on NbOX$_2$ is essential not only for potential applications but also for probing exotic quantum phenomena.

In this work, we systematically investigate HHG in bulk and flake NbOI$_2$ crystals, focusing on harmonic efficiency and polarization characteristics. Driven by intense $\sim 4~\mu\mathrm{m}$ femtosecond laser pulses, NbOI$_2$ exhibits strong, coherent broadband high-harmonic emission, comprising both odd and even orders up to at least the 16th harmonic. We further find that the harmonics display distinctly different intensity and polarization behaviors, underscoring the pivotal role of crystallographic anisotropy in governing the nonlinear optical processes in NbOI$_2$. First-principles calculations suggest that the flat band related to  Peierls dimerization plays a crucial role in enabling efficient HHG. Our results provide new pathways for materials design toward ultrafast optoelectronic and attosecond photonics applications.

\section*{Results}
\subsection*{Structural Characterization}
High-quality rectangular NbOI$_2$ single crystals with typical dimensions of $\sim$ 8 mm$\times$1 mm$\times$0.1 mm (Fig.~S1a) were synthesized using a standard chemical vapor transport (see Methods for details). The presence of several sharp $(2L00)$ diffraction peaks observed in the X-ray diffraction pattern confirms the high crystalline quality of the NbOI$_2$ crystals (Fig.~S1b). Bulk NbOI$_2$ adopts a monoclinic structure (space group No.~5, $C2$) and forms van der Waals layers stacked along the $a$ axis, as illustrated in Fig.~1a. Each layer consists of NbO$_2$I$_4$ octahedral cages, with O corners and I--I bonds oriented along the $b$ and $c$ axes, respectively. The in-plane structure is highly anisotropic, featuring one-dimensional dimerized Nb-Nb chains along the nonpolar direction. The  Nb cations confined within the octahedra exhibit a large off-center polar displacement of approximately 0.14~\AA\ along the polar $b$ axis, giving rise to a spontaneous in-plane ferroelectric polarization. Notably, the spontaneous polarization $P_s$ of NbOI$_2$ has been reported to be one of the highest values achieved in two-dimensional ferroelectric materials\cite{Jia2019,Wu2022,Wang2025}. Moreover, Fig.~1b   displays the unique in-plane network structure that resembles a square Lieb lattice, in which geometric frustration ensures electron confinement in a well-defined flat band\cite{https://doi.org/10.48550/arxiv.2510.15080}. 

\subsection*{Spectra and Intensity Dependence of HHG}
We employ an intense mid-infrared laser generated via difference-frequency generation to investigate high-harmonic generation in NbOI$_2$. Before the measurements, the bulk NbOI$_2$ crystal was freshly cleaved using mechanical exfoliation. The high-harmonic spectra were obtained from bulk NbOI$_2$ driven by linearly polarized laser pulses with a central wavelength of $4~\mu\mathrm{m}$ ($\sim 0.31~\mathrm{eV}$) at a repetition rate of 1 kHz. Intense mid-infrared femtosecond pulses were focused onto the sample, and the generated harmonics were detected in a transmission geometry (Fig.~1c).

Figure~2a presents typical HHG spectra polarized along the polar and nonpolar axes of the crystal (labeled in Fig.~1b) on a logarithmic intensity scale. The peak intensity on the sample was estimated to be $\sim 0.4~\mathrm{TW/cm^2}$, and the sample thickness was approximately $30~\mu\mathrm{m}$. The appearance of even-order harmonics directly reflects the broken global inversion symmetry of NbOI$_2$. For both driving orientations, odd- and even-order harmonics are generated over a broad spectral range from the near-infrared ($\sim$ 1.2 eV, 4th) to the deep-ultraviolet ($\sim$ 5 eV, 16th). The arrows in Fig.~2a indicate the highest detectable harmonic order within our measurement range. The emitted high harmonics significantly exceed the sample's indirect bandgap (E$_g$ =2.24 eV)\cite{Fang2021,Abdelwahab2022}, as indicated by the dashed line, and no fluorescence signal was detected in the spectra. The overall harmonic intensity is notably higher than that of a $300~\mu\mathrm{m}$ $(11\bar{2}0)$ ZnO plate under the same conditions, as displayed in Fig.~S5. The pump threshold for generating the 16th-order harmonic can be as low as $\sim 0.4~\mathrm{TW/cm^2}$, which is significantly lower than values reported in most previous HHG studies on solids (a comparison is summarized in Table~S1). For comparison, HHG spectra obtained under $2~\mu\mathrm{m}$ excitation in the transmission geometry are provided in Fig.~S7.

Next, we measured the harmonic yield as a function of pump intensity while keeping the laser linearly polarized along the principal lattice axes of the sample (Fig.~2b, c). At low pump intensity, some low-order harmonics exhibit perturbative-like scaling behavior, where the yield of the $q$th harmonic scales as $I^{q}$ (dashed lines indicate the $q$th-harmonic order). In contrast, as the pump intensity increases, the harmonic yield for all orders exhibits a cubic dependence ($I^{3}$, solid lines) on the driving laser intensity, similar to the nonperturbative scaling behavior reported in previous HHG studies on MoS$_2$, graphene, and WP$_2$\cite{Liu2017,Yoshikawa2017,Lv2021}. This behavior establishes the nonperturbative nature of the high-order generation process in NbOI$_2$.  This distinct scaling behavior indicates that the HHG observed in NbOI$_2$ originates from strong-field electron dynamics rather than conventional perturbative nonlinear processes.

\subsection*{Thickness-Dependent HHG and Related Conversion Efficiency}
In the aforementioned bulk measurements, we observed relatively strong HHG signals even at low excitation intensities. To facilitate a quantitative comparison of signal strengths, we measured the power of the 4th order harmonic and calibrated the generation efficiency for each harmonic order. An output of 59~nW was obtained for the 4th harmonic in the polar direction at a pump power of 2.7~mW, corresponding to a power conversion efficiency of $\sim 2.2 \times 10^{-5}$, demonstrating the excellent nonlinear conversion capability of the NbOI$_2$ crystal. The conversion efficiencies of the 4th--6th harmonics reach values as high as $10^{-5}$--$10^{-6}$, which are relatively large compared to those reported for other solid-state materials.
Table S2 provides a detailed comparison of HHG conversion efficiencies across various materials, along with the underlying HHG mechanisms. 

Because of the weak van der Waals interlayer coupling, thin NbOI$_2$ films can be readily obtained via mechanical exfoliation, offering unique advantages for thickness tunability. Given that both the HHG yield and cutoff order are highly sensitive to sample thickness\cite{Liu2017} and are crucial for potential nanophotonic applications, we carried out a systematic investigation of NbOI$_2$ samples with varying thicknesses. NbOI$_2$ crystals exhibit broken in-plane inversion symmetry; therefore, the generation of both even- and odd-order harmonics is expected, irrespective of the layer count. For exfoliated samples, thin NbOI$_2$ crystals were supported on the entrance plane of a $c$-cut sapphire substrate with a thickness of $300~\mu\mathrm{m}$. The bare sapphire substrate was examined and showed no detectable HHG at the low excitation fluences used in this work. 

Figure~3 shows the harmonic spectra and conversion efficiencies measured for samples with three representative thicknesses. Both bulk and thin-film NbOI$_2$ exhibit pronounced harmonic emission from the 4th to the 16th order. While the bulk sample yields broader harmonic peaks than its thinner counterparts, the overall spectral profiles remain similar, especially for the nonpolar orientation (Fig.~3a). For the polar orientation (Fig.~3b), the low-order harmonics (4th--7th) below the bandgap show significant variation in the thinner samples. The calibrated conversion efficiency results are presented in Fig.~3c–e. Notably, the higher-order conversion efficiency is enhanced for the 150-nm-thick sample, which may be attributed to reduced self-absorption in thinner layers. Harmonic radiation was compared between samples with thicknesses ranging from a few hundred nanometers to tens of micrometers under identical experimental conditions, comparable to those used for the bulk sample. The detailed thickness dependence of each high-harmonic is provided in Fig.~S6. Most harmonic signals maintain relatively consistent intensities across the entire thickness range, underscoring the strong potential for nanophotonics applications. Certain low-order harmonic orders exhibit larger oscillations, possibly due to interference effects. Previous research predicts that harmonic intensity scales quadratically below the coherence length and penetration depth, oscillates beyond the coherence length where interference effects become significant, and then saturates at larger thicknesses\cite{Guo2023,Tang_2024}. For broader harmonic orders, factors such as wavelength-dependent absorption in HHG, wavelength-dependent refractive index, and wavelength-dependent coherence length contribute to increased spectral complexity. A significant increase in harmonic efficiency by orders of magnitude was observed for low-order harmonics, potentially attributable to a longer coherence length and reduced absorption compared to high orders.

\subsection*{Dependence of HHG on Crystallographic Orientation}
The low lattice symmetry of NbOI$_2$ leads to anisotropic nonlinear optical responses, underscoring the necessity of tracing the HHG response across different crystallographic orientations. Accordingly, the HHG spectra were recorded while systematically varying the orientation angle of the linearly polarized $4~\mu\mathrm{m}$ driving laser. To avoid substrate influence, a bare thick sample was suspended over a hole for polarization-dependent measurements. In Fig.~4a, the vertical axis represents the spectrally resolved harmonic emission, and the horizontal axis displays 45 spectra arranged side by side, each corresponding to a specific polarization angle of the driving laser. The mid-infrared polarization angle was incremented in $6^\circ$ steps, allowing for comprehensive characterization of the angular dependence of the HHG spectra. The driving polarization orientation is defined relative to the sample polar axis, with $0^\circ$ corresponding to alignment along the polar axis and $90^\circ$ corresponding to the nonpolar axis.

Analysis of the measured high-harmonic spectra reveals a clear two-fold symmetric pattern in the emitted radiation. This periodicity reflects the intrinsic $C_2$ rotational symmetry of NbOI$_2$. The observed symmetry results from the interaction between the multicycle inversion-symmetric laser pulse and the crystal structure. To quantify the orientation dependence, we analyzed the spectrogram by extracting angular slices for each harmonic order, as illustrated in Fig.~4b. The harmonic signal is periodically enhanced when the driving field aligns with either the polar or nonpolar direction. 

We further investigated the HHG polarization while scanning the driving polarization orientation over $180^\circ$. The results are presented in Fig.~4c. The horizontal axis indicates the driving polarization orientation, defined as in Fig.~4a. The vertical axis represents the HHG polarization orientation which is also referenced to the polar axis. In Fig.~4c, each spectrogram corresponds to a specific harmonic order. All observed harmonics exhibit predominantly linear polarization. For even-order harmonics, the polarization is consistently aligned with the polar axis, independent of the harmonic order and the driving orientation, similar to the even-order harmonics in transition-metal dichalcogenides \cite{Kobayashi_2021}. This observation confirms that spatial inversion symmetry is broken along the polar axis, enabling exclusive generation of even-order harmonics in that direction. In contrast, odd-order harmonics display different behavior. Their polarization remains fixed to the crystal axis as the driving orientation varies between about \(\mp 45^\circ\) from the axis. When the driving polarization switches between crystal axes, the polarization of the odd-order harmonics also switches. This observation suggests that electron–hole trajectories in NbOI$_2$ predominantly follow the crystal axes, and that odd-order harmonic emission results from competition between these axes. In contrast to WS$_2$ \cite{Kobayashi_2021}, where the polarization of odd-order harmonics tracks the driving polarization, the driving polarization only mediates competition between the two crystal axes for odd-order harmonic generation in NbOI$_2$. The results can be summarized as follows: even-order harmonics are polarized along the polar axis, while odd-order harmonics are polarized along the axis that is closer to the driving orientation. The distinct polarization behaviors of even- and odd-order harmonics as functions of the driving polarization orientation suggest that they originate from different underlying HHG mechanisms. Corresponding measurements with a 2~µm driving laser are included in the SI and demonstrate similar behavior.

\subsection*{First-principles Calculations on Electronic Structures and HHG Characteristics}
Previous studies have shown that the efficiency of HHG in solids is closely related to the electronic structure. For example, efficient HHG has been observed in several topological materials\cite{ Shi2023,Bai_2020,Kovalev2020,Schmid2021,Lv2021,Heide2022}. However, systematic explorations beyond these material systems are still lacking. In particular, the relationship between correlated flat bands and HHG efficiency has not yet been well explored. Existing examples remain limited and not fully ideal, including experimental observations in CdTe and theoretical predictions in phosphorene~\cite{chen2019,Long2023}, whereas bulk NbOI$_2$ exhibits an ideal flat band across the entire Brillouin zone(BZ). To obtain a comprehensive understanding of the HHG mechanism, we perform band-structure calculations and numerically simulate the HHG spectra of NbOI$_2$ driven by mid-infrared fields. Both monolayer and bulk band structures were calculated and are presented in Fig.~5 and Fig.~S9, respectively; their overall characteristics are largely similar. Therefore, for simplicity, the discussion in the main text focuses on the monolayer NbOI$_2$.

For an isolated Nb$^{4+}$ ($4d^1$) cation, the unpaired $d$ electron half-fills the lowest-energy $d_{z^2}$ orbital due to crystal field effects, dominating the states near the Fermi level ($E_f$). Itinerant electrons would make the system metallic in the absence of lattice distortion. However, the Peierls dimerization of Nb-I-Nb chains splits the Nb $d_{z^2}$ manifold into bonding and antibonding states away from $E_f$ and leads to the full occupation of the bonding state, \cite{Ye2023,https://doi.org/10.48550/arxiv.2510.15080,Mohebpour_2024}, rendering the ground state insulating. Figure~5a shows the HSE band structure of monolayer NbOI$_2$, where $\Gamma$–X and $\Gamma$–Y denote the nonpolar and polar directions, respectively. The calculations reveal a narrow flat valence band composed of Nb $d_{z^{2}}$ orbitals, consistent with previous theoretical work\cite{Ye2023,https://doi.org/10.48550/arxiv.2510.15080}. As shown by the projected density of states in Fig.~5b, the Peierls dimerization gives rise to highly localized Nb $d_{z^2}$ orbital states. The flat band in NbOX$_2$ has been well confirmed by recent angle-resolved photoemission spectroscopy experiments\cite{https://doi.org/10.48550/arxiv.2510.15080,Bao2026}. Such a momentum-independent flat dispersion naturally leads to a high density of states, enhancing interband transitions.

Solid HHG is commonly described by two mechanisms involving intraband and interband contributions. In a conventional semiconductor, the interband mechanism is generally considered dominant, particularly for above-band-gap harmonics\cite{PhysRevLett.113.073901,Wu2015,Lanin2017}. For NbOI$_2$, a flat band exists next to the $E_f$. Due to the significant quenching of the kinetic energy, the intraband motion of carriers within the flat band is substantially suppressed, making the intraband contribution negligible in the flat band \cite{Li2025}. As shown in Fig.~5a, b, two additional I-derived valence bands also lie near the $E_f$, but their occupied density of states is lower than that of the flat band. Consequently, the interband transitions are predominantly governed by valence electrons in the flat band. The simulated HHG spectra of NbOI$_2$ along the polar and nonpolar directions, obtained by solving the Semiconductor Bloch Equations  with parameters identical to those used in the HHG experiment (See Methods), are shown in Fig.~5c, d, respectively. In the calculations, we explicitly include the three valence bands and  six conduction bands. Our numerically simulated spectra successfully reproduce the experimental results. To further distinguish the contributions to the HHG response, a simulation was conducted in which high-harmonic signals were separated into interband and intraband contributions. The results, presented in  Fig.~5e, f, indicate that for both directions, the intraband contribution is limited to the low odd-order harmonics, while the interband contribution dominates all even-order harmonics and higher odd-order harmonics. In conjunction with the experimental spectra (Fig.~1a), which show that the overall spectral features closely resemble the simulated interband contribution, these results further support the interpretation of interband enhancement arising from the flat band. 

\section*{Conclusions}
In conclusion, we experimentally demonstrate efficient and anisotropic high-harmonic generation from the two-dimensional van der Waals ferroelectric NbOI$_2$. Notably,  the overall harmonic conversion efficiency is exceptionally high compared to previously reported solid-state materials at low excitation intensities. This enhanced efficiency is closely associated with the presence of a unique correlated flat band related to the Peierls transition, which may provide guidance for searching for other efficient HHG material systems. Even-order harmonics are consistently polarized along the crystal polar axis, while the orientation of odd-order harmonics—along polar or nonpolar axes—is determined by the polarization of the driving laser field. 

Due to their readily tunable dimensionality and intrinsic ferroelectricity, polar layered transition-metal oxide halides offer unprecedented opportunities for nanoscale device applications while enabling effective control of their HHG through multiple degrees of freedom, such as light, strain, and electric-field \cite{Abdelwahab2022}. Our results demonstrate that ferroelectric van der Waals materials provide a versatile and highly tunable platform for solid-state HHG, bridging strong-field physics with emergent low-dimensional materials. This advancement opens new avenues for compact quantum light sources and nonlinear nanophotonic applications.

\section*{Methods}
\subsection*{Sample Preparation and Characterization}
High-quality NbOI$_2$ single crystals were synthesized by conventional chemical vapor transport.
To obtain the stoichiometric composition of NbOI$_2$, Nb powder, Nb$_2$O$_5$ powder, and I$_2$ pellets (3~g in total) were mixed in a molar ratio of 3:1:5 in an Ar-filled glovebox.The mixture was loaded into an 18-cm-long quartz ampoule, which was subsequently evacuated and sealed. Single crystals were grown following a previously reported procedure under identical growth conditions\cite{Handa2025}. The obtained crystals were flat and exhibited a black luster (Fig.~S1a).
To prevent possible effects from oxygen and moisture, the crystals were stored in an Ar-filled glovebox before use.

Powder X-ray diffraction measurements were carried out at room temperature using a Bruker D8 VENTURE diffractometer equipped with Cu--K$\alpha$ radiation ($\lambda = 1.5418$~\AA), spanning a $2\theta$ range of $5^\circ$--$90^\circ$. The correct stoichiometry was further verified by energy-dispersive X-ray spectroscopy. The as-grown crystal surface was cleaved on both sides before measurements, ensuring high-quality optical properties. Thin NbOI$_2$ films were fabricated by mechanical exfoliation onto 300-$\mu$m-thick \textit{c}-plane sapphire substrates. No harmonics were observed from the substrate alone at the intensities used in the experiments. The sample thicknesses used in the HHG experiments ranged from 150~nm to 30 $\mu$m, as determined using a commercial P-6 stylus profilometer (KLA Corporation). The polar and nonpolar directions of the thin films were determined by polarization-dependent second-harmonic generation measurements using an 800~nm laser in transmission geometry.
\subsection*{Experimental Setup} 
The Ti: Sapphire laser system used in this work generates 45~fs laser pulses at a central wavelength of 800~nm and a repetition rate of 1~kHz.
This output pumps an optical parametric amplifier, which produces idler pulses at $\sim$2~$\mu$m and signal pulses at $\sim$1.33~$\mu$m.
These beams are directed into a  mid-infrared difference-frequency generation  stage, generating an output near 4~$\mu$m in a 1 mm thick AgGaS$_2$ crystal.
A series of filters is employed to isolate the mid-infrared beam from residual signal and idler light. Two wire-grid polarizers are placed in the optical path to ensure high polarization purity. The first polarizer is also used to adjust the mid-infrared beam intensity, while the second remains fixed. The polarization of the mid-infrared beam is set by a 4~$\mu$m MgF$_2$ half-wave plate. After passing through these components, the mid-infrared beam is focused onto the sample surface using a ZnSe lens. The $1/e^{2}$ intensity radius at focus is approximately $120~\mu\mathrm{m}$, smaller than the sample dimensions. High-harmonic emission from the sample is focused into a 600-$\mu$m-diameter optical fiber through two CaF$_2$ lenses, then the fiber delivers the signal to thermoelectrically cooled spectrometers (Ideaoptics). An optional wire-grid polarizer placed between the sample and the CaF$_2$ lens is used to analyze the polarization state of the emitted harmonics. All HHG spectra are measured in a transmission geometry and under ambient conditions. Additional details of the optical setup, including all optical components, are provided in Fig.~S4.

The high HHG efficiency enables direct measurement of the harmonic flux using a calibrated photodiode power sensor (S120C, Thorlabs), which is insensitive to the mid-infrared driving laser. To minimize the influence of stray light, a long-pass filter (FELH0900, Thorlabs) and a short-pass filter (FESH1100, Thorlabs) were placed immediately before the sensor to isolate the 4th harmonic from the $30~\mu\mathrm{m}$-thick NbOI$_2$ sample. The mid-infrared power was monitored with a thermal sensor (S401C, Thorlabs). The conversion efficiency is defined as $\eta_{4th} = P_{\mathrm{4th}} / P_{\mathrm{MIR}}$, where $P_{\mathrm{4th}}$ and $P_{\mathrm{MIR}}$ denote the measuring power of the 4th harmonic and the mid-infrared driving field, respectively. Conversion efficiencies for the remaining harmonic orders were extracted from the measured harmonic spectra.

\subsection*{First-principles Calculations on Electronic Structures and HHG Characteristics} 
First-principles calculations of two-dimensional NbOI$_2$ were performed within density functional theory (DFT) as implemented in the Vienna \textit{ab initio} Simulation Package (VASP)~\cite{Kresse_1996}. 
The generalized gradient approximation (GGA) in the Perdew--Burke--Ernzerhof (PBE) parametrization and projector augmented-wave (PAW) pseudopotentials were employed~\cite{Bl_chl_1994,Kresse_1999}. 
The plane-wave energy cutoff was set to 500~eV, and a vacuum layer of 20~\AA{} was introduced to avoid spurious interactions between periodic images. 
The Brillouin zone was sampled using $6\times12\times1$ and $6\times12\times6$ $k$-point meshes for the monolayer and bulk phases, respectively. 
All geometric structures were fully relaxed until the convergence criteria for the total energy and atomic forces were less than $10^{-6}$~eV and $10^{-2}$~eV/\AA{}, respectively. 
To correct the PBE-calculated bandgap closer to the experimentally measured one, additional electronic-structure calculations were performed using the Heyd--Scuseria--Ernzerhof (HSE06) hybrid functional~\cite{Heyd2006,Brunetto_2012}. 
Further computational details regarding the semiconductor Bloch equations (SBEs) and DFT calculations are provided in Table.~S3-6.  

We investigated high-harmonic generation (HHG) in two-dimensional NbOI$_2$ under intense laser fields by solving the extended multiband semiconductor Bloch equations (SBEs)~\cite{McDonald_2015,Jiang_2018}. 
Three conduction bands and six valence bands are included to accurately capture the electronic-structure contributions to HHG. 
We consider a linearly polarized laser propagating along the optical axis of the crystal, with the initial valence bands fully occupied by electrons. 
Consistent with previous studies~\cite{Luu_2016}, the calculations are restricted to the laser polarization direction. The time-dependent intraband electric current and interband polarization are calculated as 
$J(t)=\sum_{\lambda,\mathbf{k}}\left[-2\,v_{\mathbf{k}}^{\lambda}f_{\mathbf{k}}^{\lambda}(t)\right]$ 
and 
$P(t)=\sum_{\lambda,\lambda',\mathbf{k}}\left[d_{\mathbf{k}}^{\lambda\lambda'}p_{\mathbf{k}}^{\lambda\lambda'}(t)+\mathrm{c.c.}\right]$, 
where $\lambda=e,h$ denotes electrons or holes, $v_{\mathbf{k}}^{\lambda}$ is the group velocity derived from the band dispersion, $f_{\mathbf{k}}^{\lambda}(t)$ is the time-dependent carrier population, $d_{\mathbf{k}}^{\lambda\lambda'}$ is the transition dipole matrix element, and $p_{\mathbf{k}}^{\lambda\lambda'}(t)$ is the microscopic interband polarization amplitude. The HHG spectrum intensity is obtained from the Fourier transform as 
$S(\omega)\propto\left|\omega P(\omega)+iJ(\omega)\right|^2$.  The pump laser is centered at a wavelength of $\lambda_0 = 3950~\mathrm{nm}$, corresponding to an optical period of $T = 13.18~\mathrm{fs}$. It delivers a peak intensity of $I_0 = 0.5 \times 10^{12}~\mathrm{W/cm^2}$, which translates to a peak electric field strength of $E_0 = 1.37 \times 10^{9}~\mathrm{V/m}$. The intensity full width at half maximum (FWHM) spans seven complete optical cycles, corresponding to a temporal duration of $\tau_{\mathrm{FWHM}} \approx 92.3~\mathrm{fs}$. The dephasing time is set to $T_2 = 0.25\,T$, where $T$ is the optical period of the pumping laser.

\section*{Data Availability}
The datasets generated and/or analyzed during the current study are available from the corresponding author upon reasonable request.

\bibliography{NbOI2}

\newpage
\section*{Acknowledgments}
This work was supported by the National Natural Science Foundation of China (Grant Nos.~12304375 and 12488201) and the National Key Research and Development Program of China (Grant Nos.~2024YFA1408701 and 2022YFA1403901).

\section*{Author Contributions}
T.H., C.Z., and N.W. conceived the study. 
T.H. and C.Z. designed and built the experimental setup and performed the optical measurements, with assistance from J.H., R.D., H.W., and Q.W. F.L. and C.Q. performed first-principles calculations under the supervision of R.L. T.H. synthesized and characterized NbOI$_2$ single crystals. Q.L. prepared  NbOI$_2$ flakes. T.H., C.Z., and N.W. analyzed and interpreted the data and wrote the manuscript, with critical input from all authors. 

\section*{Competing Interests}
The authors declare no competing interests.

\newpage

\begin{center}
\begin{figure}
   \sbox0{\includegraphics[width=\textwidth]{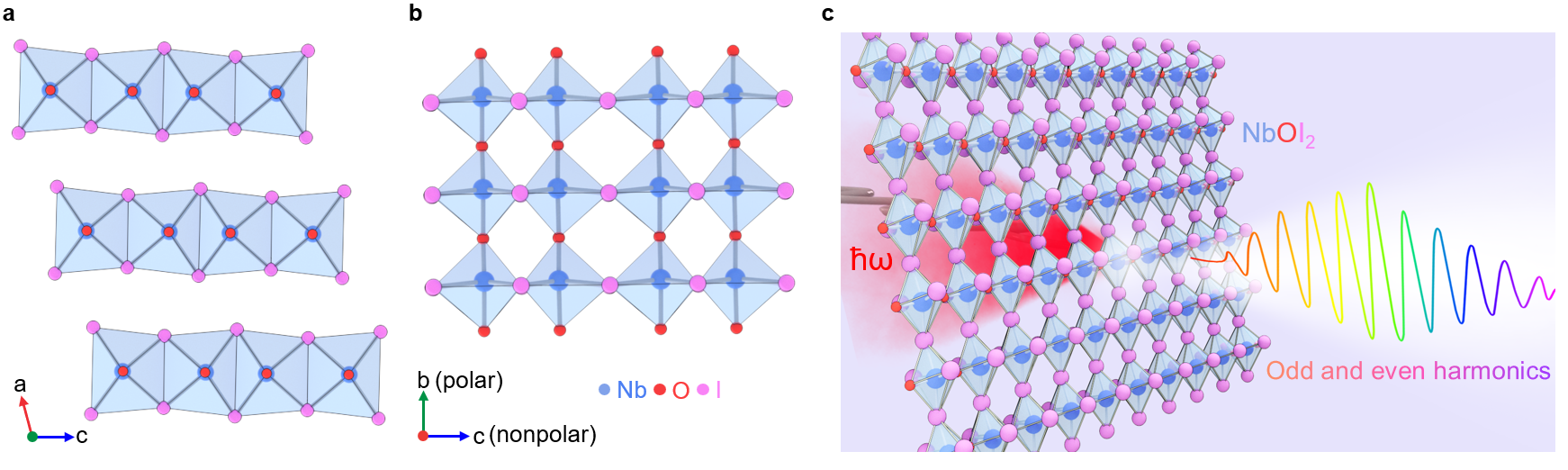}}
    \begin{minipage}{\wd0}
  \usebox0
  \captionsetup{labelfont={bf},labelformat={default},labelsep=period,name={Fig.}}
  \caption{\textbf{Schematic atomic structure of NbOI$_2$ and  illustration of the experimental setup.}  \textbf{a} Front view of the crystal structure of NbOI$_2$ (space group, $C2$), Nb (blue), O (red), and I
(pink) atoms form quasi-2D layers stacked along the a axis. 
\textbf{b} Top view of NbOI$_2$ along the $a$ axis. The crystal possesses in-plane intrinsic ferroelectricity, such that the $b$ and $c$ directions correspond to the polar and nonpolar directions, respectively. Nb-Nb atoms along the $c$ axis undergo Peierls dimerization, resulting in a $2 \times 1$ superstructure. The lattice geometry projected onto the bc plane reveals a network closely resembling a Lieb lattice. \textbf{c} Schematic illustration of mid-infrared transmission HHG spectroscopy in NbOI$_2$, with odd- and even-order harmonics generated and collected from the back surface of the crystal.
\label{fig:expsetup}}
\end{minipage}
\end{figure}  
\end{center}

\begin{center}
\begin{figure}
   \sbox0{\includegraphics[width=\textwidth]{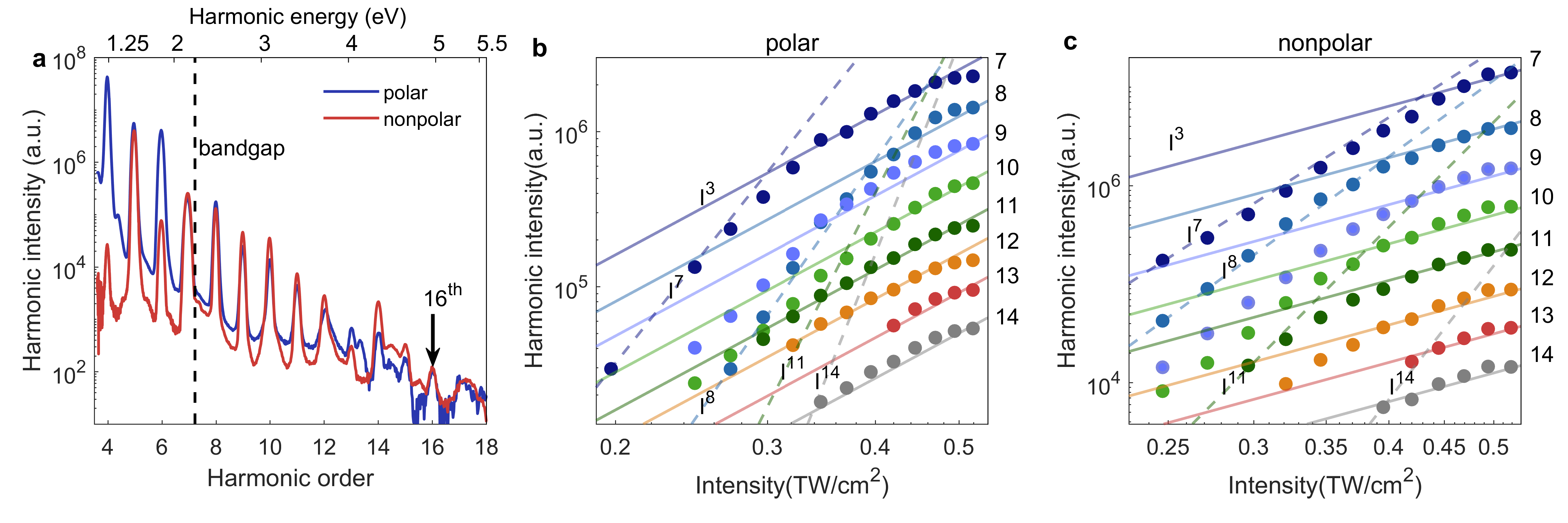}}
    \begin{minipage}{\wd0}
  \usebox0
  \captionsetup{labelfont={bf},labelformat={default},labelsep=period,name={Fig.}}
  \caption{\textbf{Measured high-harmonic generation (HHG) spectra, intensity dependence of different harmonic orders from a bulk NbOI$_2$ crystal.
} 
\textbf{a} High-harmonic spectra from bulk NbOI$_2$ driven at a central wavelength of 4~$\mu$m. Blue/red curves correspond to the driving beam oriented along the polar/nonpolar axis, respectively. 
\textbf{b,c} Measured individual harmonic yield as a function of peak pump intensity $I$ (dots), with the driving laser oriented along the  polar($b$) and nonpolar($c$)  axes of the sample, respectively. Experimental data are fitted based on the power law $I^{p}$ (solid lines). The behavior strongly deviates from that expected for a perturbative nonlinear response with $p=q$ (dashed lines, scaled to the experimental data at the lowest intensity).
\label{fig:faradayCD}}
\end{minipage}
\end{figure}  
\end{center}
\begin{center}
\begin{figure}
   \sbox0{\includegraphics[width=\textwidth]{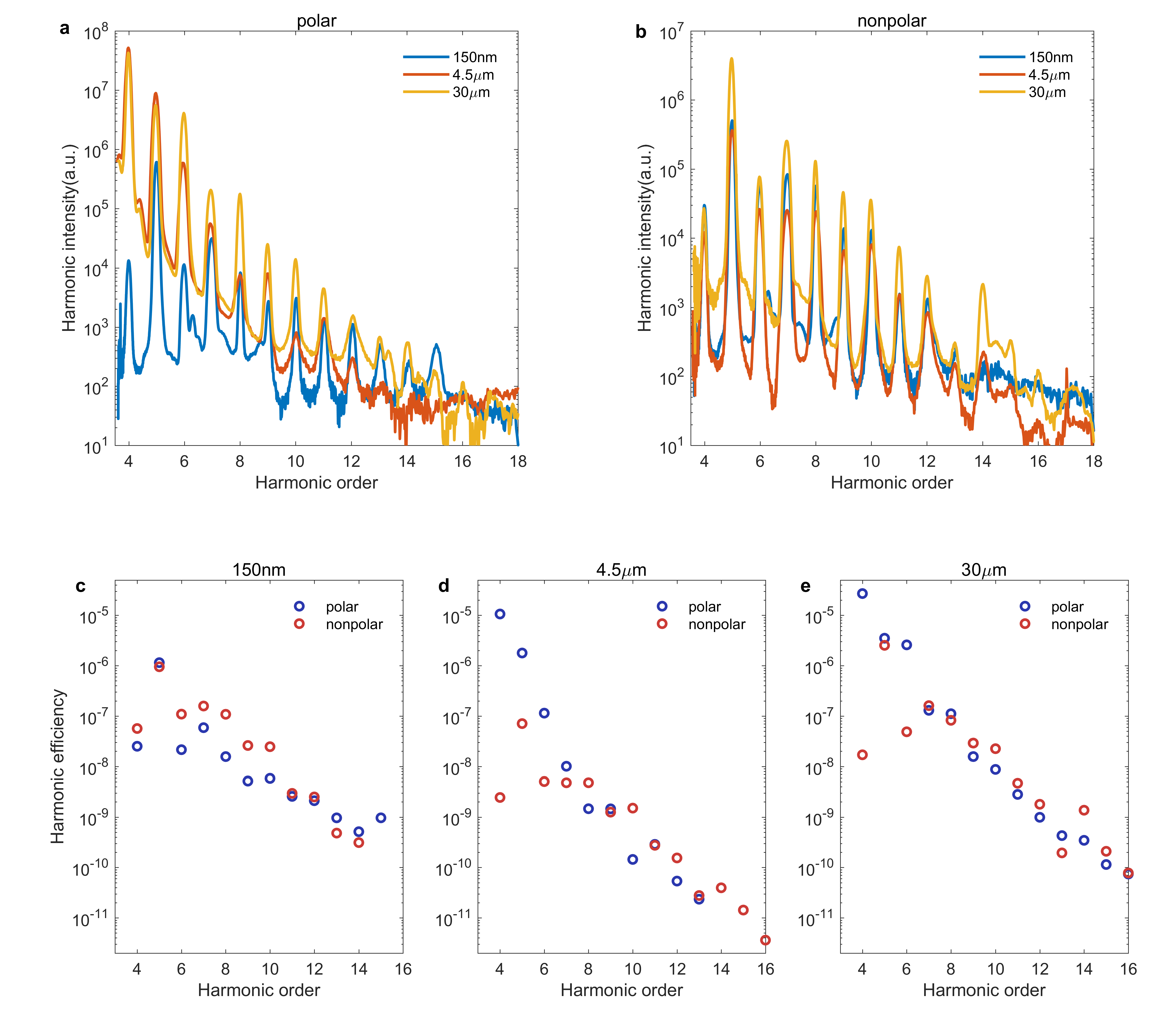}}
    \begin{minipage}{\wd0}
  \usebox0
  \captionsetup{labelfont={bf},labelformat={default},labelsep=period,name={Fig.}}
  \caption{\textbf{Measured HHG spectra and related harmonic efficiency from three different thicknesses of NbOI$_2$.
} 
High-harmonic spectra obtained from NbOI$_2$ samples of varying thicknesses with the driving laser oriented along the \textbf{a} polar and \textbf{b} nonpolar axes of the sample, respectively. 
\textbf{c}--\textbf{e} Harmonic generation efficiency for each harmonic order from 4th to 16th for samples with different thicknesses.
\label{fig:faradayCD}}
\end{minipage}
\end{figure}  
\end{center}

\begin{center}
\begin{figure}
   \sbox0{\includegraphics[width=\textwidth]{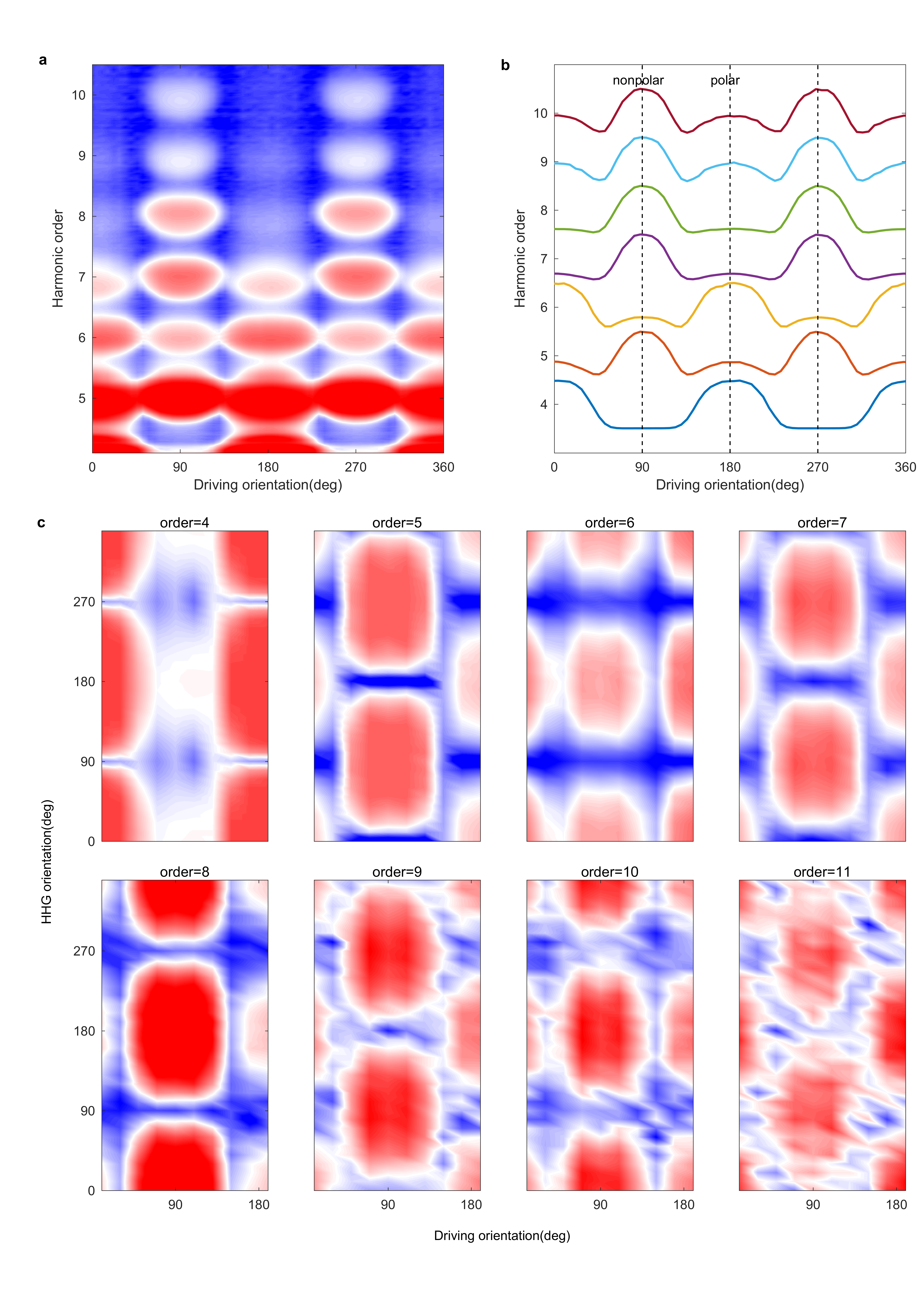}}
   \begin{minipage}{\textwidth}
     \centering
     \scalebox{0.8}{\usebox0}  
     \captionsetup{labelfont={bf},labelformat={default},labelsep=period,name={Fig.}}
     \caption{\textbf{High-harmonics intensity from a NbOI$_2$ crystal as a function of the driving orientation and the generating harmonics orientation.}
    \textbf{a} HHG spectra as a function of the orientation angle of the laser polarization relative to the polar optic axis in a 30~$\mu$m-thick NbOI$_2$ sample driven at 4~$\mu$m. The harmonics exhibit maxima (and minima) every 90$^\circ$ of rotation. Intensity is shown in false color on a logarithmic scale (arbitrary units). 
\textbf{b} Normalized harmonic intensities for the perpendicular configuration, obtained by integrating the corresponding spectral regions. 
\textbf{c} HHG polarization measurements as a function of the orientation angle of the laser polarization relative to the polar optic axis. Each spectrogram corresponds to a specific harmonic order, as labelled in the figure. Intensity is shown in false color on a logarithmic scale (arbitrary units). 0$^\circ$ corresponds to the polar axis of the sample.}
\label{fig:faradayCD}
   \end{minipage}
\end{figure}
\end{center}

\begin{center}
\begin{figure}
   \sbox0{\includegraphics[width=\textwidth]{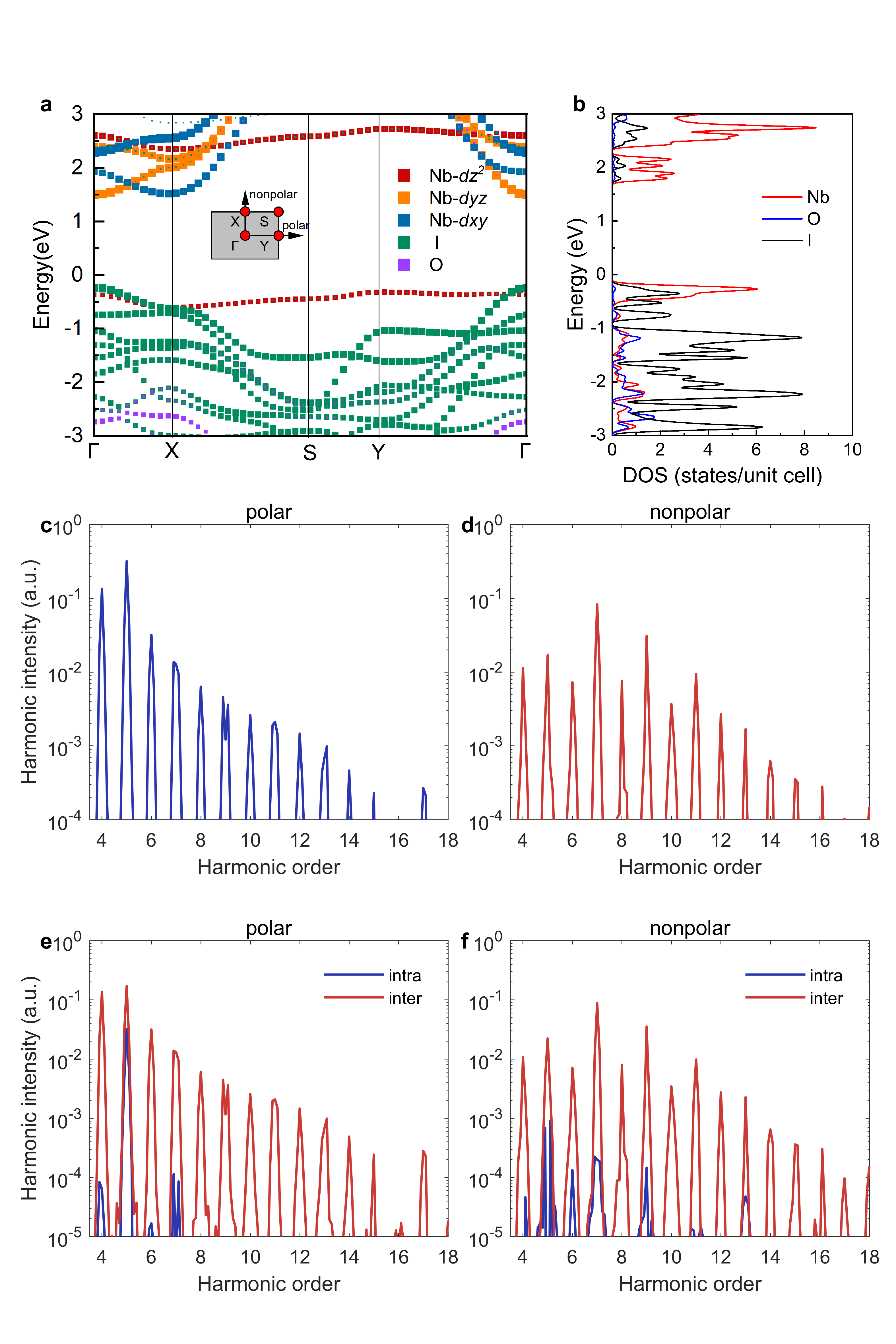}}
   \begin{minipage}{\textwidth}
     \centering
     \scalebox{0.8}{\usebox0}  
     \captionsetup{labelfont={bf},labelformat={default},labelsep=period,name={Fig.}}
     \caption{\textbf{Calculated Band Structure and HHG Spectra of NbOI$_2$.}
    \textbf{a}  HSE-calculated band structure of monolayer NbOI$_2$ with orbital projection. A flat band predominantly derived from Nb bonding states near $E_f$ is separated by a gap from higher conduction bands. \textbf{b} The related density of states of NbOI$_2$, which reveals a pronounced peak from the Nb orbital near the $E_f$.
\textbf{c, d} The numerically calculated HHG spectra along the polar/nonpolar directions, respectively. The  HHG spectra were separated into interband and intraband contributions for driving orientation along polar direction in \textbf{e} and nonpolar direction in \textbf{f}, respectively.}
\label{fig:faradayCD}
   \end{minipage}
\end{figure}
\end{center}

\end{document}